# Virtual Black Holes in Hyperbolic Metamaterials

Igor I. Smolyaninov

*Department of Electrical and Computer Engineering, University of Maryland, College Park, MD 20742, USA*

**Optical space in electromagnetic metamaterials may be engineered to emulate various exotic space-time geometries. However, these metamaterial models are limited in many respects. It is believed that real physical space-time strongly fluctuates on the Planck scale. These fluctuations are usually described as virtual black holes. Static metamaterial models introduced so far do not exhibit similar behavior. Here we demonstrate that thermal fluctuations of optical space in hyperbolic metamaterials lead to creation of virtual electromagnetic black holes. This effect is very large if the dielectric component of the metamaterial exhibits critical opalescence.**

Recent breakthroughs in nanofabrication enabled rapid progress in development of novel electromagnetic metamaterials [1], which gain their properties from structure rather than composition. Considerably extended range of material parameters (local dielectric permittivity $\varepsilon_{ik}$ and magnetic permeability $\mu_{ik}$ tensors) in these media enabled novel approaches to electromagnetic design, which take advantage of the field theoretical ideas developed to describe physics in curvilinear space-times [2,3]. Recent examples of unusual "optical spaces" designed based on this "transformation optics" approach include electromagnetic black holes [4-8], wormholes [9], rotating cosmic strings [10], and the Alcubierre warp drive [11]. It was also demonstrated that phase transitions in electromagnetic metamaterials are capable of emulating physical processes which took place during and immediately after the Big Bang [12,13]. These models can be very informative for phenomena where researchers have no direct experience and therefore limited intuition. However, the metamaterial models introduced so far capture only static classic physics features of these unusual space-time metrics. Space time fluctuations, which become very strong on the Planck scale $(\hbar G/c^3)^{1/2}$ have not been emulated.

While exact physical behaviour of space-time on the Planck scale is not yet known, uncertainty principle gives us some important insights. If we try to probe any space-time event with the Planck scale precision, we would need to concentrate so much energy that the Planck-size space-time region would become a black hole. Such virtual black holes are believed to disrupt the continuity of space-time on the Planck scale [14]. Here we consider hyperbolic metamaterials [15,16] made using dielectrics which exhibit critical opalescence [17], and demonstrate that "optical space" in such metamaterials also exhibits strong fluctuations. Moreover, similar to the Planck scale physics, these fluctuations give rise to virtual electromagnetic black holes. Thus, using hyperbolic metamaterials we may emulate (to some degree) microscopic properties of space-time on the Planck scale.

Let us consider a metal-dielectric composite hyperbolic metamaterial (Fig.1) in which the dielectric component exhibits critical opalescence. Critical opalescence is very easy to observe in mixtures of fluids, such as cyclohexane and aniline. However, it was also reported in such solid dielectrics as quartz in the vicinity of the α-β transition [18]. Aniline and cyclohexane are immiscible below $T_c$=35 ºC. In this immiscible region there exists an aniline-rich and a cyclohexane-rich phase, which have large refractive index difference: $n_{anilin} \sim 1.6$ while $n_{cyclohexane} \sim 1.4$. On the other hand, above $T_c$ the liquids mix well with each other. Near the critical point the domains present in the mixture can switch easily between aniline-rich and cyclohexane-rich phases. These domains demonstrate fractal shapes on the scale of many micrometers. Hence, light is scattered strongly by the mixture across a small temperature range around the transition temperature: the effective dielectric permittivity seen by light experiences wide fluctuations in space and time in the $1.96 < \varepsilon_d < 2.56$ range. At the critical point, the scattering is so intense that the system becomes opaque.

Let us consider the effective dielectric permittivity tensor of such a hyperbolic metamaterial following calculations in ref. [19]. We will consider a simpler case of "layered" hyperbolic metamaterial (see Fig.1(b)). The diagonal components of the permittivity tensor for extraordinary photons in this case are obtained as

$$\varepsilon_{x,y} = n_m \varepsilon_m + (1-n_m)\varepsilon_d \qquad (1)$$

$$\varepsilon_z = \frac{\varepsilon_m \varepsilon_d}{(1-n_m)\varepsilon_m + n_m \varepsilon_d} \quad , \qquad (2)$$

where $n_m$ and $\varepsilon_m$ are the volume fraction and the dielectric constant of the metal component, respectively. (In the case of "wired" metamaterials the results are qualitatively similar: eq.(1) gives an approximate result for $\varepsilon_z$, while eq.(2) gives $\varepsilon_{x,y}$). Since $\varepsilon_m$ is negative and frequency dependent, a "hyperbolic frequency band" (in which $\varepsilon_{x,y} < 0$, while $\varepsilon_z > 0$) may appear in the dispersion law of the extraordinary photons for a given $n_m$ [19]. Hyperbolic metamaterials were experimentally realized in [20,21]. Let us assume for simplicity that $n_m \ll 1$ and that the metal component is described by the Drude model:

$$\varepsilon_m = 1 - \frac{\omega_p^2}{\omega^2 + i\omega\gamma}, \tag{3}$$

where $\gamma$ is small. The Drude model parameters of such good metal as silver are $\omega_p$=1.37x10$^{16}$ s$^{-1}$ and $\gamma$=7.29x10$^{13}$ s$^{-1}$ [22]. As a result, eqs.(1,2) become

$$\varepsilon_{x,y} \approx \varepsilon_d - \frac{n_m \omega_p^2}{\omega^2 + i\omega\gamma} \quad , \quad \varepsilon_z \approx \varepsilon_d \tag{4}$$

and the metamaterial medium is hyperbolic at

$$\omega < \omega_0 = \left(\frac{n_m}{\varepsilon_d}\right)^{1/2} \omega_p \quad , \tag{5}$$

Let us demonstrate that in the lossless continuous medium limit fluctuations of the "optical space" (due to fluctuations of $\varepsilon_d$) look like virtual electromagnetic black holes. Just below the boundary $\omega_0$ of the hyperbolic frequency band $\varepsilon_{x,y}$ is negative and very close to zero. The dispersion law of extraordinary photons can be written as [15,16]

$$k_x^2 + k_y^2 = \frac{\varepsilon_z \omega^2}{c^2} - \frac{\varepsilon_z k_z^2}{\varepsilon_{x,y}} \tag{6}$$

Therefore, in the limit $\varepsilon_{x,y} \to -0$ we obtain $k^2 \approx k_x^2 + k_y^2 \approx -\varepsilon_z k_z^2 / \varepsilon_{x,y}$. The "optical length" element experienced by the extraordinary photons equals

$$dl_{opt}^2 = \frac{k^2 c^2 dl^2}{\omega^2} \approx -\frac{\varepsilon_z k_z^2 c^2 dl^2}{\omega^2 \varepsilon_{x,y}} \quad , \tag{7}$$

where $dl$ is the length element in the coordinate space. Let us consider an extraordinary photon propagating towards a region where due to fluctuations of $\varepsilon_d$ instantaneous $\varepsilon_{x,y} \to -0$. As an example, we can assume a local instantaneous distribution of $\varepsilon_{x,y}$ of the form $\varepsilon_{x,y} = -\alpha x^2$ where $\alpha>0$. Corresponding "optical length" element is

$$dl_{opt}^2 \approx \frac{\varepsilon_z k_z^2 c^2 dl^2}{\omega^2 \alpha x^2} \qquad (8)$$

This length element coincides with the length element as perceived by a Rindler observer near a black hole event horizon. The Rindler metric can be written as [23]

$$ds^2 = -\frac{g^2 x^2}{c^2} dt^2 + dl^2 \quad, \qquad (9)$$

The horizon is located at $x=0$. The spatial line element of the corresponding Fermat metric as perceived by the Rindler observers is

$$dl^2 = \frac{dl^2}{-g_{00}} = \frac{c^4 dl^2}{g^2 x^2} \qquad (10)$$

Comparison of eqs.(8) and (10) demonstrates that a region of fluctuating "optical space" near $x=0$ looks instantaneously like a virtual electromagnetic black hole.

As a result, eqs.(6-10) indicate that probing "optical space" on large scales (using very small $k_z \rightarrow 0$) would produce some average dielectric permittivity of the fluctuating medium. On the other hand, probing the "optical space" on smaller scales would result in observation of microscopic virtual black holes. This behaviour is rather similar to the behaviour of actual physical space-time on the Planck scale. Appearance of the virtual microscopic black holes is illustrated in Fig.2. Fig.2(a) shows a randomly generated instantaneous microscopic fractal map of dielectric permittivity in a mixture of aniline and cyclohexane just below the critical temperature $T_c$. Fig.2(b) shows how this map looks if probed with small $k_z$ photons. Fig.2(c) demonstrates a similar fluctuating permittivity distribution, as it would be perceived by extraordinary photons propagating in a hyperbolic metamaterial at $\omega$ just below $\omega_0$. Areas shown in red appear as "virtual black holes". Qualitatively, this result can be understood as follows. In the limit $\varepsilon_{x,y} \rightarrow -0$ the extraordinary photon momentum parallel to the boundaries of the red areas diverges: $k_x^2 + k_y^2 \to \infty$. This means that photons cannot escape these areas: they experience total internal reflection at any incidence angle.

In conclusion, we have considered hyperbolic metamaterials made using dielectrics exhibiting critical opalescence, and demonstrated that effective "optical space" in such materials may experience strong fluctuations. Similar to the space-time behaviour on the Planck scale, these fluctuations give rise to virtual electromagnetic black holes. Thus, using hyperbolic metamaterials we may emulate such an interesting feature of the real physical space-time as virtual microscopic black holes. This effect appears to be large and easy to observe.

**Figure Captions**

**Figure 1.** Schematic views of the (a) "wired" and (b) "layered" nonlinear hyperbolic metamaterials made of either metal wires or metal layers inside a dielectric Kerr medium. (c) Hyperbolic dispersion relation allowing unbounded values of the wavevector (blue arrow).

**Figure 2.** (a) Randomly generated instantaneous microscopic fractal map of dielectric permittivity in a mixture of aniline and cyclohexane just below the critical temperature $T_c$. (b) Same map as it looks if probed with small $k_z$ photons. (c) A similar fluctuating permittivity distribution, as it would be perceived by extraordinary photons propagating in a layered hyperbolic metamaterial at $\omega$ just below $\omega_0$. Areas shown in red appear as "virtual black holes".

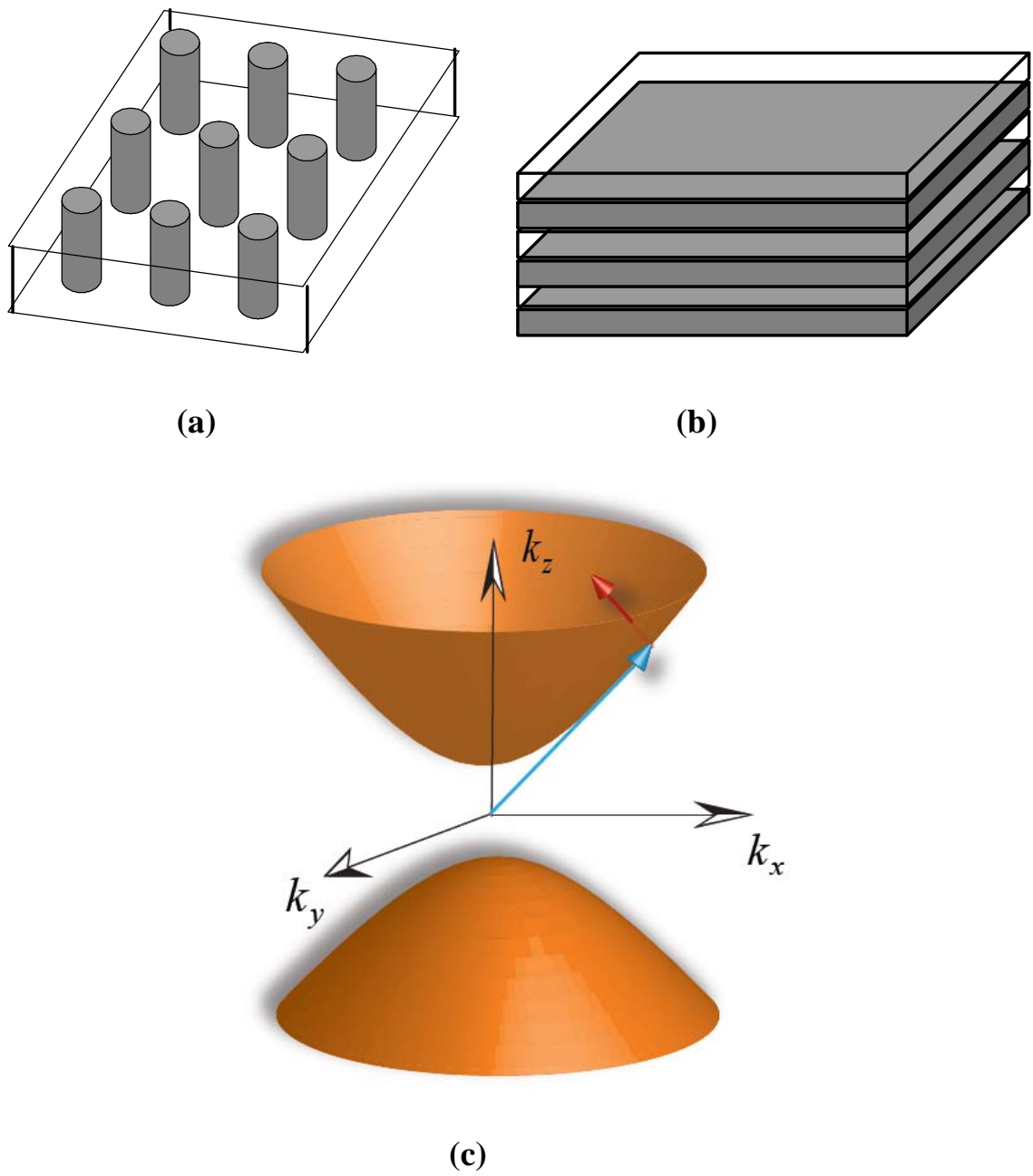

**Fig.1**

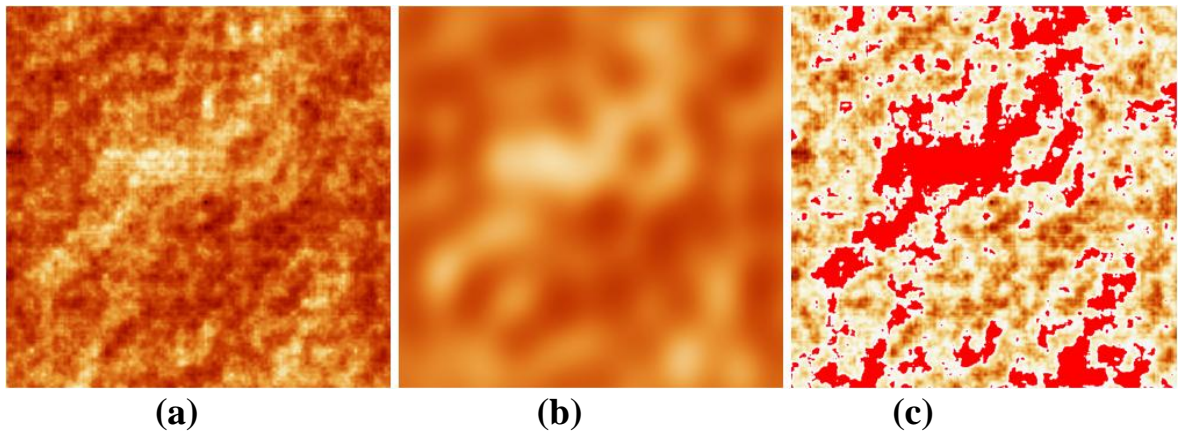

**(a)** **(b)** **(c)**

**Fig.2**